\documentstyle[preprint,pre,aps,eqsecnum]{revtex}
\input epsf.sty
\begin{document}
\begin{titlepage}

\title{Analysis of Optimal Velocity Model with Explicit Delay}
\author{
	Masako Bando\thanks{e-mail address: bando@aichi-u.ac.jp},\
	Katsuya Hasebe\thanks{e-mail address: hasebe@aichi-u.ac.jp} \\ 
	{\normalsize\em Physics Division, Aichi University,
		Miyoshi, Aichi 470-02, Japan}
	\vspace{10pt} \\
	Ken Nakanishi\thanks{e-mail address:
		kenichi@yukawa.kyoto-u.ac.jp} \\
	{\normalsize\em Department of Physics,
		Nagoya University,
		Nagoya 464-0814, Japan}
	\vspace{10pt} \\
	Akihiro Nakayama\thanks{e-mail address:
		g44153g@nucc.cc.nagoya-u.ac.jp} \\
	{\normalsize\em Gifu Keizai University, Ohgaki, Gifu 503-8550, Japan}
}
\maketitle

\begin{abstract}
We analyze Optimal Velocity Model (OVM) with explicit delay.
The properties of congestion and the delay time of car motion 
are investigated by analytical and numerical methods. 
It is shown that the small explicit delay time has almost no effects.
In the case of the large explicit delay time, a new phase of congestion 
pattern of OVM seems to appear. 
\end{abstract}

\vspace{1cm}



\end{titlepage}

\section{Introduction}

In recent years, we proposed a new car-following model 
``Optimal Velocity Model'' (OVM) based on a dynamical equation \cite{OVM1}
\begin{equation}
\ddot{x}_n(t)=a \left\{ V(x_{n+1}(t)-x_n(t))-\dot{x}_n(t) \right\}\ ,
\label{neweq}
\end{equation}
where $t$ is time and $x_n$ is a position of the $n$-th car.
Cars are numbered so that the $(n+1)$-th car precedes the $n$-th car.
The driver feels the headway $x_{n+1}(t)-x_n(t)$ and determines
an optimal velocity $V(x_{n+1}(t)-x_n(t))$.
It is best to drive a car with the optimal velocity but
in general a deviation exists between the optimal velocity and a real one.
The driver responds to the deviation
$\Delta V =V(x_{n+1}(t)-x_n(t)) -\dot{x}_n(t)$ and diminishes it 
by giving an acceleration $a\Delta V$ to the car.
The coefficient $a$ expresses the sensitivity of the driver.
We call the function $V$ ``Optimal Velocity Function'' (OVF).
In previous papers, we have shown how OVM can explain 
behaviors of traffic flow, for example, the transition from a free flow
to a congested flow, a density-flow relationship, a kind of 
effective delay of car motion\cite{OVM1,OVM2,OVM3,OVM4}. 

On the other hand, the prototype equation of motion of traditional 
car-following model is
\begin{equation}
\ddot{x}_n=\lambda_0 \{ \dot{x}_{n+1}-\dot{x}_n \}\ ,
\label{proto}
\end{equation}
where $\lambda_0$ is a constant\cite{Pipes,Gazis,Newell}.
In this model, a driver is thought to react to the stimulus proportional to
the relative velocity between the previous car and his own car.
Equation (\ref{proto}) may be generalized by changing the constant $\lambda_0$
to a function $\lambda(x_{n+1}-x_n)$ of headway.
However these models have no physically interesting solution because
such equation can be integrated easily and be reduced to following equation
\begin{equation}
\dot{x}_n = V(x_{n+1}-x_n)\ ,
\label{nextproto}
\end{equation}
where $V$ is a function of headway and 
$V'(x_{n+1}-x_n) = \lambda(x_{n+1}-x_n)$. In car-following models,
therefore, the introduction of ``delay'' is necessary and 
plays an essential role to understand the traffic 
dynamics \cite{Kometani,Text}.
Following equation is a typical one which is widely used in car-following
models.
\begin{equation}
\dot{x}_n(t+\tau)=V(x_{n+1}(t)-x_n(t)),
\label{old}
\end{equation}
where $\tau$ is a delay time of driver's response.
The driver senses headway at time $t$ and changes
the velocity of his car at later time $t+\tau$.
This delay time $\tau$ of response has been thought to be inevitable 
because it comes from the driver's physical delay of response to the stimulus 
together with the mechanical response time of a car.
In this paper, this $\tau$ will be called ``explicit delay time''.

The notion of explicit delay time $\tau$ is completely different 
from that of ``delay time of car motion'' 
introduced in our previous paper \cite{OVM4}. 
Let us recall the definition of the delay time of car motion.
Consider a pair of cars, a leader and a follower.
Assume the leader changes the velocity according to $v_l=v_0(t)$ and
the follower duplicates the leader's velocity 
but with some delay time $T$, that is, $v_f=v_0(t-T)$.
Under such a situation we can clearly define the delay time of 
car motion by $T$. It is known that
the observed delay time $T$ of car motion is of the order
of 1 sec, but the known physical or mechanical response 
time $\tau$ is of the order of 0.1 sec.
In the previous paper we confirmed that 
the equation (\ref{neweq}) really produces $T$ of order 1 sec.

We clarified that OVM can describe the properties of traffic flows or
the behaviors of cars fairly well without any explicit delay time $\tau$. 
However there exists the delay time of response of driver for a fact. 
The explicit delay time $\tau$ should be included in the dynamical equation 
in order to construct realistic models of traffic flow. 
It is a natural question what kind of effect appears 
in the traffic flow or in the car motion if we introduce the explicit delay 
in the equation(\ref{neweq}). 

In this paper we investigate the following equation
\begin{equation}
\ddot{x}_n(t+\tau)=a\left\{ V(x_{n+1}(t)-x_n(t))-\dot{x}_n(t)\right\}\ .
\label{new}
\end{equation}
In order to our analysis be more concrete, we use the parameter
$a=2.0$ (1/sec) and the function $V$ which are phenomenologically determined
in previous papers by the observed data on Japanese motorways
\cite{Koshi,Oba,Xing}.
\begin{equation}
V(\Delta x) = 16.8\left[
	\tanh{0.0860\left(\Delta x - 25\right)} + 0.913\right]\ ,
\label{OVreal}
\end{equation}
in which the unit of length and time are 'meter' and 'second' respectively.

The plan of this paper is as follows. In section 2
we discuss the global properties of traffic flow in OVM 
with the explicit delay. In section 3 we investigate 
more microscopic property, that is, the delay time of 
car motion. First we discuss within a linear approximation and next 
evaluate the delay times of car motion in various cases by 
numerical simulations. As a special case, the car motion 
under the traffic signal is also treated. 
In section 4 we show the new feature of OVM with the explicit delay.
The final section is devoted to summary and discussion.

\section{Property of Traffic Flow in OVM with Explicit Delay}

\subsection{Linear Analysis}

In this section we investigate OVM with the explicit delay time $\tau$ of 
driver's response described by equation (\ref{new}).

First we analyze the linear stability of $N$-car system
on a circular lane of length $L$.
Obviously, the homogeneous flow solution of equation (\ref{new}) is given by
\begin{equation}
x_n^{(0)}(t)=V(b)t+nb, \ \ \ \ \ b=L/N\ \ .
\label{eq:uniformsol}
\end{equation}
To see whether the solution (\ref{eq:uniformsol}) is stable or not,
we add a small perturbation 
\begin{equation}
x_n(t) = x_n^{(0)}(t) + y_n(t)\ .
\label{eq:perturb}
\end{equation}
From equation (\ref{eq:perturb}) and equation (\ref{new}),
we can calculate a linearized equation with respect to $y_n(t)$
\begin{equation}
\ddot y_n(t+\tau) = a\{ f\Delta y_n(t)-\dot y_n(t)\}, 
\label{eq:lineareq}
\end{equation}
where $f = V'(b)$ and $\Delta y_n=y_{n+1}-y_n$.
A complete set of solutions is given by
\begin{equation}
y_{jn}(t)= \exp (i\alpha_j n + i\omega_j t ),
\label{eq:linearsol} 
\end{equation}
where $\alpha_j = 2\pi j/N$ for $j = 1,2,3,\cdots,N$ and
$\omega_j$ satisfies the equation
\begin{equation}
-\left(\frac{\omega_j}{a}\right)^2
 \exp \left[i\left(\frac{\omega_j}{a} \right)a \tau \right]
= \left(\frac{f}{a}\right) \left\{ e^{i\alpha_j} - 1 \right\}
 -i\left(\frac{\omega_j}{a}\right) \ .
\label{eq:omega}
\end{equation}
In equation (\ref{eq:omega}), variables are combined to be dimensionless.
The condition that each solution $y_{jn}(t)$ becomes marginally stable 
is Im~$\omega_j = 0$. For convenience of explanation, we will omit the 
mode-index $j$ and treat $\alpha$ as a continuous variable.
The condition Im~$\omega = 0$ gives `critical curves' for each $a\tau$ 
in $(f/a,\ \alpha)$ plane, where $f/a$ is a radial coordinate and $\alpha$
is an angular coordinate.
Mode solutions $y_{jn}(t)$ are represented by a point $(f/a, \alpha_j)$
on a circle $f/a$=const.

\begin{figure}[ht]
\epsfysize=8cm
\hspace*{2cm}
\epsfbox{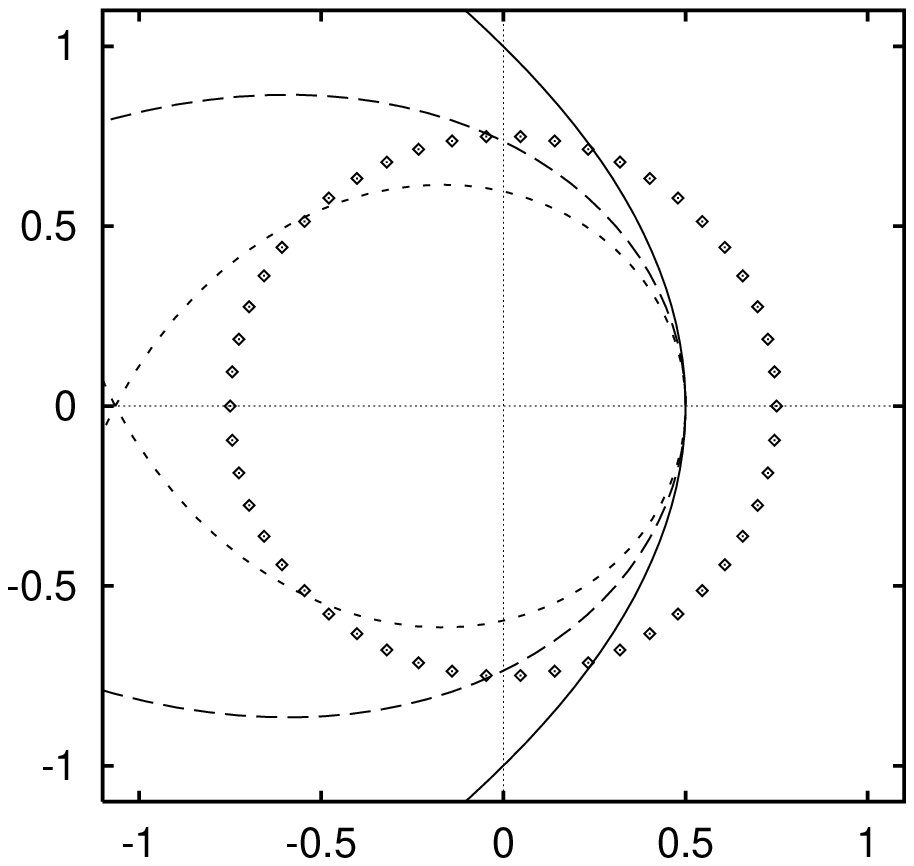}
\caption{Solid line, dashed line and dotted line show critical curves 
	for $a\tau$ = 0, 0.2 and 0.4, respectively. A circle of 
	diamond marks represents mode solutions for $f/a = 0.75$.}
\label{Fig1}
\end{figure}

Three critical curves for $a\tau$ = 0, 0.2 and 0.4 are shown in 
figure \ref{Fig1}, in which a reference circle represents mode solutions 
for $f/a = 0.75$. The modes staying outside (right-hand side) of the
critical curve are unstable. Figure \ref{Fig1} shows that a homogeneous 
flow state with a parameter $f/a = 0.75$ is an unstable state.
From figure \ref{Fig1}, it is found that unstable modes
increase as the explicit delay time $\tau$ becomes large.
This situation looks similar to a case the sensitivity $a$
becomes small in the original OVM\cite{OVM1}. 
Though there seems to be some relationship between the sensitivity $a$ 
and the explicit delay time $\tau$ as indicated in Ref.\cite{Nagel},
an analytical relation has not been clarified yet.

\subsection{Numerical Simulations}

\begin{figure}[ht]
\epsfysize=8cm
\hspace*{2cm}
\epsfbox{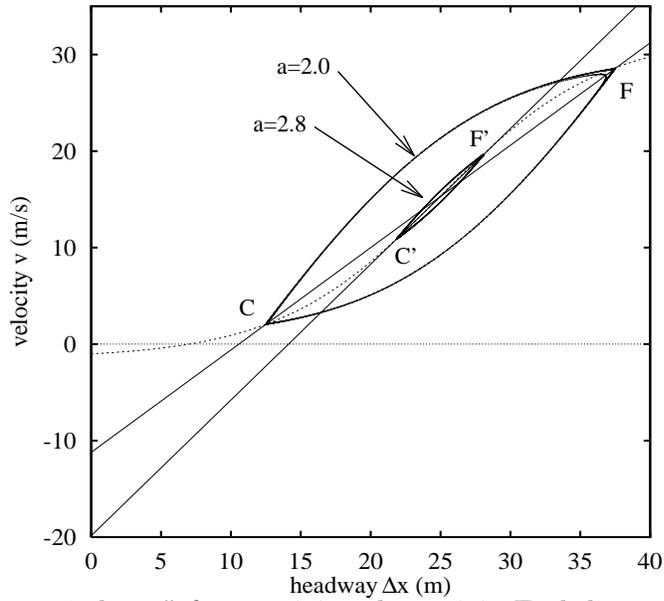}
\caption{``Hysteresis loops'' for $a=2.0$ and $a=2.8$.
	Each line connects two turning points of each hysteresis loop.
	A tanh-type curve represents OVF (\ref{OVreal}).}
\label{Fig2}
\end{figure}

The effect of the explicit delay in the congestion formation
can be evaluated by numerical simulations. 
In previous papers\cite{OVM1,OVM2}, we investigated the property of 
traffic flows in a circuit.
It is found that when the car density exceeds a critical value, 
a homogeneous traffic flow becomes unstable and makes a phase transition 
to a congested flow. After enough time, the congested flow becomes stationary
and shows an alternating pattern of high density (congestion cluster) and 
low density regions.
Each velocity and headway inside high (low) density regions 
always take common values which are determined only by the sensitivity $a$
and OVF independently of any other conditions. 
The motion of each car can be shown in a `phase space' ($\Delta x$, $\dot x$),
and the trajectories draw a single ``hysteresis loop'', a kind of limit cycle.
Figure \ref{Fig2} shows typical hysteresis loops for sensitivity $a=2.0$
and $2.8$. Two turning points $C=(\Delta x_c,v_c)$ and $F=(\Delta x_f,v_f)$ 
correspond to the high and low density region for $a=2.0$ and 
$C'$ and $F'$ for $a=2.8$. We found the congestion pattern 
moves backward on the circuit with a constant velocity 
$(v_f \Delta x_c - v_c \Delta x_f)/(\Delta x_f - \Delta x_c)$, 
which is given by 
the intersection of $\dot x$ axis and the line connecting two turning points 
$C$ and $F$. Therefore the property of such congested flows is almost decided 
by two points $C$ and $F$ of hysteresis loop.

\begin{figure}[ht]
\epsfysize=8cm
\hspace*{2cm}
\epsfbox{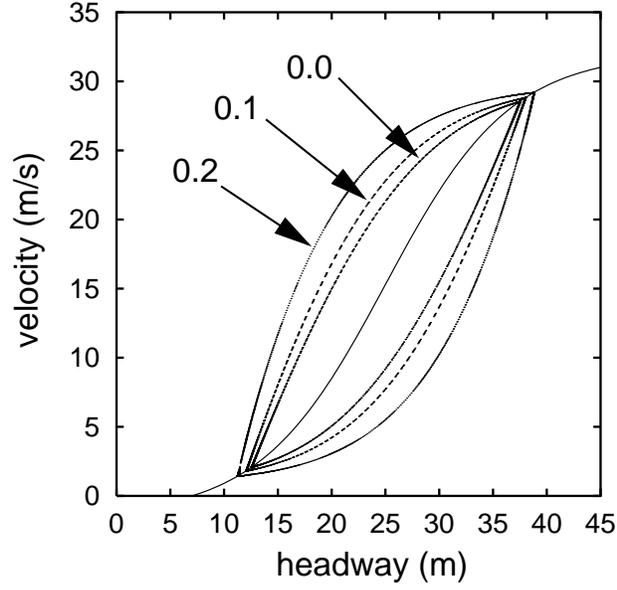}
\caption{Hysteresis loops for $\tau$ = 0, 0.1 and 0.2. 
	A tanh-type curve represents OVF (\ref{OVreal}).}
\label{Fig3}
\end{figure}

From numerical simulations, we recognize no qualitative difference 
in the behavior of the traffic flow 
between the cases with and without the explicit delay, 
if $\tau$ is not so large.
Figure \ref{Fig3} shows hysteresis loops for $\tau$ = 0, 0.1 and 0.2,
that is, $a\tau$ = 0, 0.2 and 0.4. 
The changes of hysteresis loops are similar to those for the case 
that the sensitivity $a$ becomes small in the original OVM\cite{OVM1}.
Therefore it seems that the explicit delay time $\tau$, which is not 
so large, does not play any essential role in the congestion formation.
In other words, the effect of the explicit delay can be 
almost compensated by the change of sensitivity $a$.

\begin{figure}[ht]
\epsfysize=8cm
\hspace*{2cm}
\epsfbox{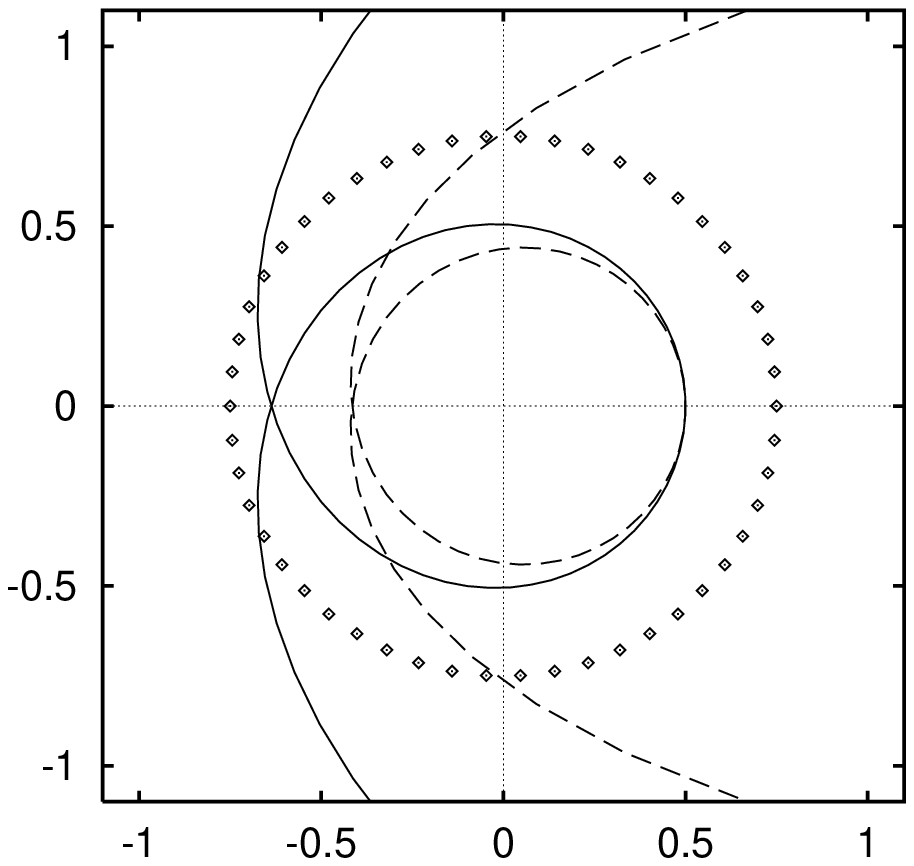}
\caption{Solid line and dashed line show critical curves 
	for $a\tau=0.6$ and $a\tau=0.8$, respectively. A circle of 
	diamond marks represents mode solutions for $f/a = 0.75$.}
\label{Fig4}
\end{figure}

Obviously, this is not the case for a very large $a\tau$.
Figure \ref{Fig4} shows examples for $a\tau = 0.6,\ 0.8$ where 
critical curves are inside the referenced circle $f/a=0.75$.
In original OVM instability always comes from 
long range modes $(\alpha \sim 0)$, that is, short range modes 
$(\alpha \sim \pi)$ are
always stable. In the case $a\tau > 0.6$, however, there exist 
various cases in which all modes become unstable or short range modes only 
become unstable. In such cases, the instability starts from 
all modes or from short range modes. 
It is interesting to see what kind of phenomena emerge in such cases.
An example shall be discussed in section 4.

\section{Delay Time of Car Motion}

\subsection{Linear Analysis}

In this section, we investigate the delay of car motion 
in order to see the effect of the explicit delay from a more 
microscopic viewpoint. First, we analyze the delay of car motion
with the linear approximation.

Consider a pair of a leader and it's follower where
the leader moves with the velocity $v(t)$ and
the follower replicates the motion of the leader
after the time interval $T$, that is, 
the follower's velocity is given by $v(t-T)$.
In this case we can define the delay time of car motion as $T$.

Let the position of the leader at time $t$
be $y(t)$ and that of it's follower $x(t)$ which obeys
equation (\ref{new}), that is,
\begin{equation}
\ddot{x}(t + \tau) = a\{ V(y(t) - x(t))-\dot{x}(t)\}\ .
\label{leader_follower}
\end{equation}
Starting from the situation with headway $b$ and velocity $V(b)$,
\begin{equation}
y_0(t)=V(b)t + b,\ \ \ x_0(t)=V(b)t\ ,
\label{eq:const_motion}
\end{equation}
we investigate the response of the follower $\xi (t)$ to 
a small change $\lambda (t)$ of the leader:
\begin{equation}
y(t)=y_0(t) + \lambda (t),\ \ \ 
x(t)=x_0(t) + \xi (t) .
\label{001}
\end{equation}
Inserting above equations 
into equation (\ref{leader_follower}) and taking a linear 
approximation, we get
\begin{equation}
\ddot{\xi}(t + \tau) + a\dot{\xi}(t) + af \xi(t) = af \lambda(t) , 
\label{002}
\end{equation}
where $f = V'(b) $ is again a derivative of OVF 
at headway $b$.
If one takes $\lambda(t) = e^{i\omega t}$,
the solution is given by
\begin{equation}
\xi(t) = \frac{1}{1 + i\omega/f - e^{i\omega \tau}\omega^2/af}
\ e^{i\omega t}\ .
\label{eq:E8}
\end{equation}
This is rewritten as
\begin{equation}
\xi(t) = |\xi|\ e^{i\omega (t - T)}\ ,
\label{eq:E9}
\end{equation}
where
\begin{eqnarray}
T &=& \frac{1}{\omega}\tan^{-1}\frac{a\omega-\omega^2\sin(\omega\tau)}
{af-\omega^2\cos(\omega \tau)}\ , \\
|\xi| &=& \left[ 1+(\frac{\omega}{f})^2 -2(\frac{\omega^2}{af})
	(\cos \omega \tau + \frac{\omega}{f}\sin\omega\tau)
	+ (\frac{\omega^2}{af})^2 \right ]^{-\frac{1}{2}}\ .
\end{eqnarray}
First let us consider the case $|\omega| $ is sufficiently small
($\omega /f \ll 1,\omega /a \ll 1 $).
It will be discussed later whether this condition is satisfied or not
in the realistic situation used in equation (\ref{OVreal}).
Then we have
\begin{equation}
|\xi| = 1,\ \ \ T = \frac{1}{f}.
\label{003}
\end{equation}

Here we take the general expression of $\lambda (t)$
which is expressed as follows.
\begin{equation}
\lambda(t) = \int \tilde {\lambda}(\omega)e^{i\omega t} d\omega \ .
\end{equation}
$\tilde {\lambda} (\omega)$ is assumed to be nonzero only for 
$\omega$ small enough.
Then we find the follower's response becomes
\begin{equation}
\xi(t)= \int \tilde {\lambda}(\omega)e^{i\omega (t - T)} d\omega \ 
= \lambda(t - T)\ ,
\end{equation}
that is,
\begin{equation}
\dot x(t) = V(b) + \dot\xi(t) = V(b) + \dot\lambda(t - T) = \dot y(t-T),
\end{equation}
with $T$ of equation(\ref{003}).

As a result we conclude that for sufficiently slow 
and small change of leader's velocity, 
the delay time $T$ of motion of the follower becomes $1/f$ 
(the inverse of derivative of OVF at corresponding headway), 
independently of the explicit delay time $\tau$ of driver's response.

\subsection{Simulations for Homogeneous Flows}

Next we will carry out numerical simulations to investigate
the effect of the explicit delay in homogeneous traffic flows. 
The validity of the conditions $\omega \ll a,\ f$ can be checked also.
We make simulations starting from homogeneous flows with various headways and
add a small disturbance to the first car. 
The delay time of car motion is estimated between 10th car and 11th car
when the disturbance propagates there.

In table \ref{Tab1}, we summarize the results
of numerical simulation. 
In the cases where the homogeneous flow is stable, 
the delay time $T$ of car motion is almost equal to $1/f$ and
the explicit delay has no effect.
The cases $\Delta x =20,\ 25,\ 30$ correspond to the unstable situation.
The measurement of the delay time $T$ is carried out before 
the disturbance becomes large. The results show that 
the assumption $\omega \ll a,\ f$ is not valid.
Even in such cases the explicit delay does not affect $T$.

\subsection{Simulation for Congested Flows}

In this subsection, we treat the car motion in a stationary congested flow,
where linear analysis is no more valid obviously.
In the previous paper\cite{OVM4} we have shown that
the delay time $T$ of car motion is the inverse of the gradient of line 
which connect two turning points ($C$ and $F$ in figure \ref{Fig2}).
This is a natural extension of the statement obtained by the 
linear analysis:
``The delay time of car motion is the inverse of derivative of
OVF at corresponding headway''.

Our task here is to carry out similar numerical simulations with
the explicit delay. After the congestion pattern becomes stable, 
all cars behave in the same manner expressed in figure \ref{Fig3}.
We can estimate the delay time $T$ from the time interval
of the motion of successive two cars, which is equivalent to 
the gradient of line 
connecting two turning points of the hysteresis loop.
Table \ref{Tab2} shows the results of simulations
for $\tau=0,\ 0.1,\ 0.2$.

The table clearly shows that the change of $T$ is rather small
compared to the change of $\tau$. Therefore 
the main contribution of the delay of car motion 
comes from the structure of OVM itself and
not from the explicit delay. The $\tau$-dependence of 
$T$ appears only through the change of turning points
of the hysteresis loop. In other words, 
the effect of the explicit delay is similar to the change of 
the sensitivity $a$ and is not essential in the same as the previous section.

\subsection{Simulations for Car Motion under a Traffic Signal}

In this subsection we study the delay of motion of cars starting
from a traffic signal. Though this may be a special case 
compared to previous subsections, experiments to observe the delay time 
have often been done in this situation. 

Numerical simulations are carried out as follows. 
First a traffic signal is red and all cars are 
waiting with headway 7 (m), at which OVF (\ref{OVreal}) becomes zero.
At time $t=0$, the signal changes to green and cars start. 

\begin{figure}[ht]
\epsfysize=8cm
\hspace*{2cm}
\epsfbox{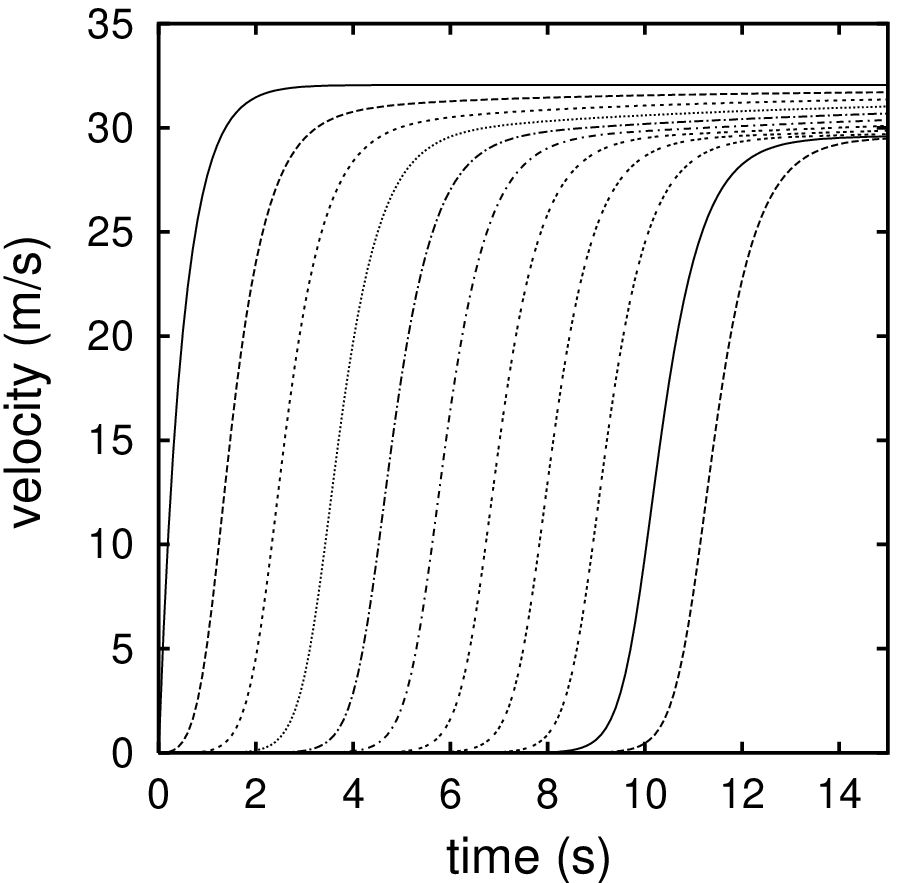}
\caption{Motions of No.1-11 cars for $\tau=0$. Each curve shows 
	the velocity of each car.}
\label{Fig5}
\end{figure}

\begin{figure}[ht]
\epsfysize=8cm
\hspace*{2cm}
\epsfbox{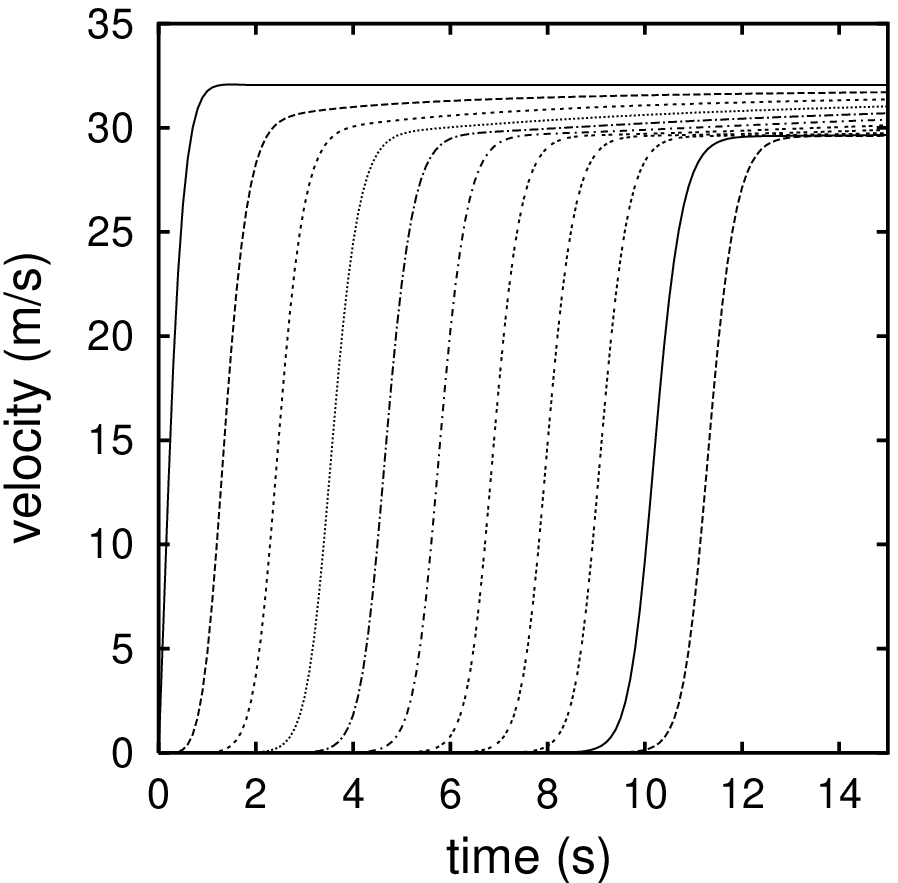}
\caption{Motions of No.1-11 cars for $\tau=0.2$. Each curve shows 
	the velocity of each car.}
\label{Fig6}
\end{figure}

Figures \ref{Fig5} and \ref{Fig6} show the velocities of several 
cars in a queue for the cases of $\tau = 0$ and $\tau = 0.2$ (sec), 
respectively. It can be seen that 
cars with large car number (7th or more) behave almost in the same manner 
as its preceding car.

We can estimate the delay time $T$ from the behavior of velocities 
of 7-10th cars. Table \ref{Tab3} shows the delay time of car motion 
for various $\tau$. Again we find that the delay time $T$ has 
a small dependence on the explicit delay time $\tau$.
To see whether this is general or not, we carried out 
another simulations with the initial headway 3 (m).
For this purpose, OVF (\ref{OVreal}) is changed 
to take zero for $\Delta x < 7$ (m). 
We show the results in the third column of table \ref{Tab3}, 
which again show obviously a small dependence of $T$ on $\tau$. 

Hitherto we concerned the definition of delay time of car motion given in the 
section 1: if velocities of two successive cars are given by $v(t)$ and 
$v(t-T)$ respectively, the delay time of car motion is $T$.
This definition is valid only for the case that the motions of two cars 
are similar. As is seen from figures \ref{Fig5} and \ref{Fig6},
the first several cars move in the different
manner, because the headway of the first car is infinite but that of other
cars are relatively small. In order to explore the delay time of car motion 
in such case we will propose an another definition.
For example, we can define the delay time as the interval 
between the time when the preceding car starts and the time 
when the next car starts. Though there are many other possibilities,
the above definition looks rather natural.

\begin{figure}[ht]
\epsfysize=8cm
\hspace*{2cm}
\epsfbox{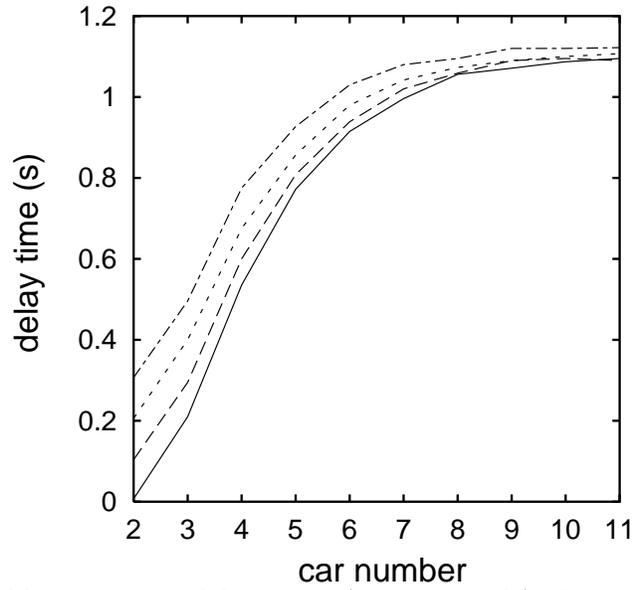}
\caption{Solid line connects delay times (time intervals) of 2-11th cars 
	for $\tau=0$. Dashed, dotted and dashed dotted lines connect
	those for $\tau=$ 0.1, 0.2 and 0.3, respectively.}
\label{Fig7}
\end{figure}

Figure \ref{Fig7} shows the delay time of car motion by the new 
definition. Obviously the data
approach to a certain values as the car number becomes large.
The limits of the delay times in this definition are the same
values as those in the previous definition. 
It should be mentioned that the explicit delay time $\tau$ is 
simply added to the delay time $T$ of car motion for first a few cars.
This effect dissipates after several cars start.

From these results, we can conclude that the explicit delay time 
$\tau$ contributes directly to the delay time of car motion 
only for such a restricted case as for the motion of first a few cars
starting from the traffic signal. In general case, the
contribution of $\tau$ to $T$ is rather small and is similar to the
contribution from the change of the sensitivity $a$.

\section{New Features of OVM with Explicit Delay}

In this section we show new features which exist only in OVM with the 
explicit delay.

\subsection{Overshoot Phenomenon}

\begin{figure}[ht]
\epsfysize=8cm
\hspace*{2cm}
\epsfbox{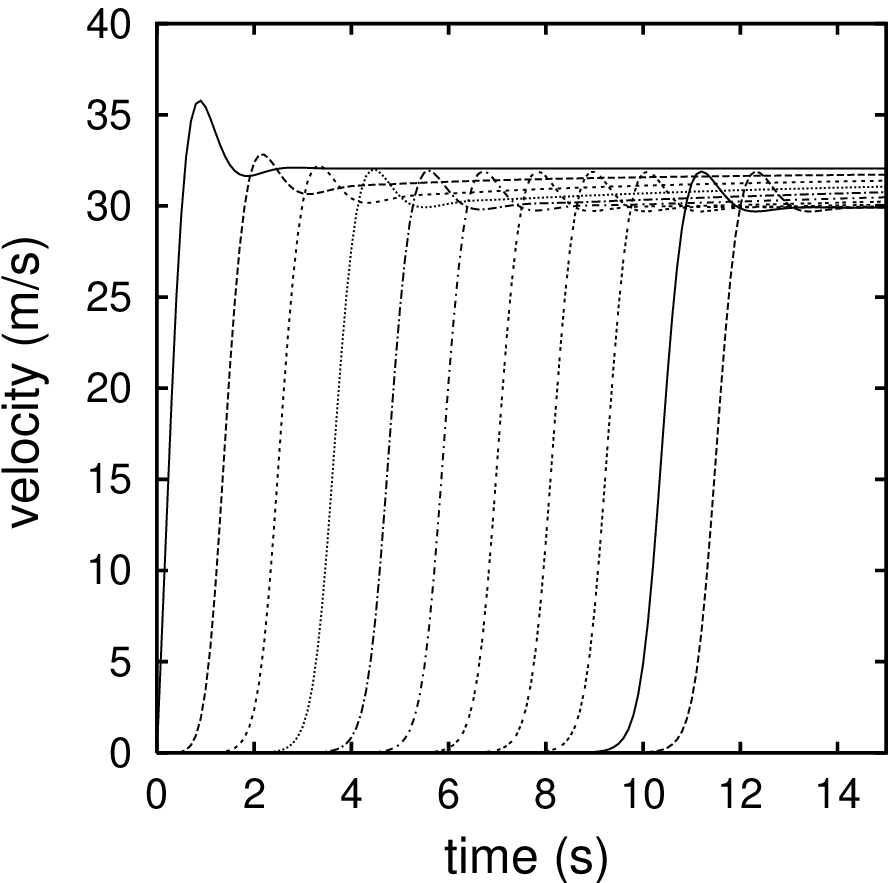}
\caption{Motions of No.1-11 cars for $\tau=0.3$. Each curve shows 
	the velocity of each car.}
\label{Fig8}
\end{figure}

We investigated the motion of cars controlled by a traffic signal 
in the previous section. For small $\tau$ the motions of cars are 
not so different from those for $\tau=0$. For large $\tau$, however, 
we can see a transitional overshoot of velocity, that is, a excess 
and a gradual decrease of velocity. As a typical case, 
the motions of cars for $\tau=0.3$ are shown in figure \ref{Fig8}.
We have carried out many numerical simulations by changing $\tau$ 
and found that the overshoot phenomenon begins at $\tau = 0.19$ (sec).

\subsection{Upper Bound of $\tau$}

First we note that the explicit delay time $\tau$ is understood
as the summarized effect coming from delays of physical and 
mechanical response. Therefore too large value will not be permitted
from observations. There exists, however, more restrictive bound,
which has a origin in the equation of motion (\ref{neweq}) of OVM.

We consider a homogeneous equation of the linearized equation
(\ref{002}) in the leader-follower system:
\begin{equation}
\ddot{\xi}(t + \tau) + a \dot{\xi}(t) + af\xi (t) = 0\ .
\label{eqn:homo}
\end{equation}
$\xi(t)$ gives a perturbative motion of the follower when the leader
moves in a constant velocity. In order that the two body 
system is stable, $\xi(t)$ must vanish as time develops.
We see that $\xi(t)=e^{i\omega t}$ is a solution of equation (\ref{eqn:homo}),
with $\omega$ satisfying
\begin{equation}
-{\omega}^2 e^{i\omega \tau}+ ia\omega + af = 0.
\label{eqn:homocond}
\end{equation}
The marginally stable condition Im $\omega$ = 0 becomes
\begin{equation}
a\tau = \kappa \sin(\kappa)\ ,\ \ f\tau = \kappa \cot(\kappa)\ ,
\label{eqn:marg}
\end{equation}
where $\kappa \equiv \hbox{Re }\omega\ \tau$. By eliminating $\kappa$,
we can find the upper bound of $\tau$ for given $a$ and $f$.
Though we could not solve equation (\ref{eqn:marg}) analytically,
the upper bound $\tau_m$ is found to be a monotonic decreasing function
of both $a$ and $f$. 

The value of $\tau_m$ can be evaluated numerically.
For the sensitivity $a$ = 2.0 (1/sec) and the maximum value 
of $f$ = 1.44 (1/sec), which is read off from OVF (\ref{OVreal}),
the corresponding upper bound $\tau_m$ is 0.44 (sec).
If $\tau > \tau_m $ (sec) in OVM with above $a$ and $f$,
the car cannot follow the constant velocity motion of the leader.
Thus $\tau_m$ should be understood as the upper bound of the explicit delay
time in order that OVM is meaningful as a model of traffic flow.

\subsection{New Congestion Pattern}

Inside the above upper bound of the explicit delay time, some curious 
phenomena emerge in traffic flow as the explicit delay time becomes large. 
If such phenomena should be regarded as unrealistic, 
the upper bound will be taken at a smaller value. 

\begin{figure}[ht]
\epsfysize=8cm
\hspace*{2cm}
\epsfbox{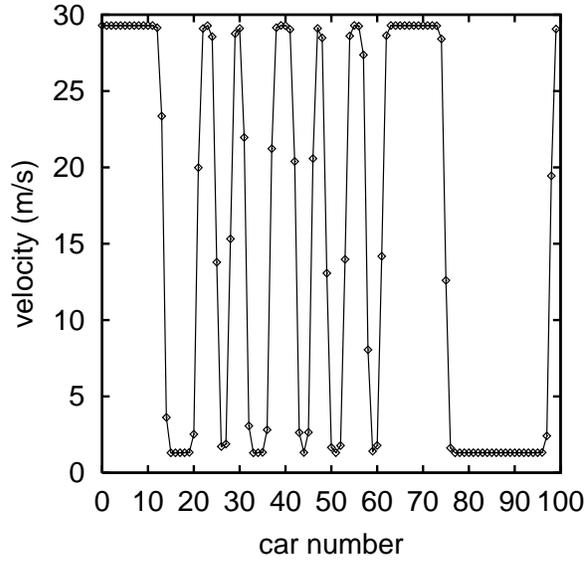}
\caption{A snapshot of velocities at $t=10000$ (sec). 
	Diamond marks represent velocities of cars. }
\label{Fig9}
\end{figure}

Figure (\ref{Fig9}) shows a snap shot of headway after enough simulation time.
The conditions of the simulation are as follows:
total car number $N$ = 100, circuit length $L$ = 2500 (m) and 
explicit delay time $\tau$ = 0.22 (sec).
There we can see small congestion clusters or rapid change of velocity 
between 15th and 60th car.
This pattern looks like a intermediate pattern before 
the congestion is formed completely. 
However, in contrast to the case of $\tau = 0.20$ (sec) where such 
a pattern or small congestions disappear as time goes, 
the pattern has very long life and may never disappear 
in the case of $\tau = 0.22$ (sec). This pattern occupies a larger region
as $\tau$ increase. 

\begin{figure}[ht]
\epsfysize=8cm
\hspace*{2cm}
\epsfbox{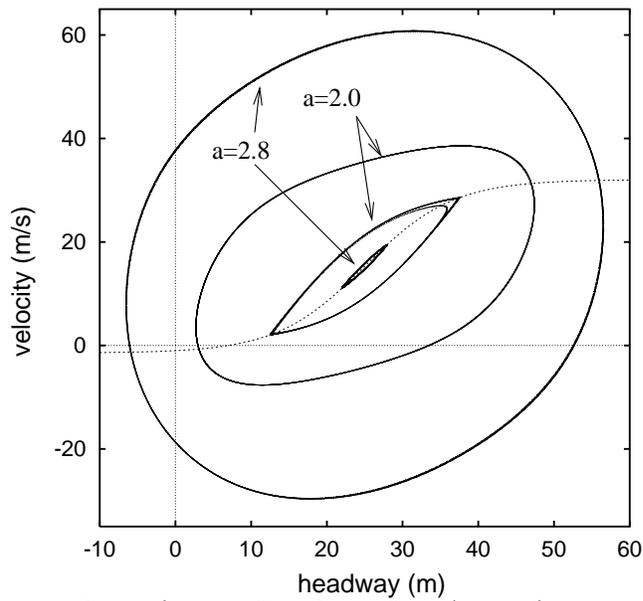}
\caption{Hysteresis loops for $\tau=0$ and $\tau=0.4$. As a reference,
	two cases, $a=2.0$ and $a=2.8$, are shown.
	A tanh-type curve represents OVF (\ref{OVreal}).}
\label{Fig10}
\end{figure}

Next we take $\tau$ to be a larger value 0.4 (sec). 
Figure \ref{Fig10} shows the hysteresis loops for $\tau = 0$ and 0.4.
Here we note that OVF (\ref{OVreal}) takes negative value continuously 
for $\Delta x < 7$ (m) and therefore cars can move backward 
(without collisions).
Because such behaviors of vehicles are obviously unrealistic,
it seems natural to set the upper bound of $\tau$ to the transition 
point at which this hysteresis loop appears.

As shown in figure \ref{Fig10},
the profiles of hysteresis loops are qualitatively different. 
Moreover the hysteresis loop for $a=2.8$ is larger than that 
for $a=2.0$ in the case of $\tau=0.4$ in contrast to the case of $\tau=0$. 
We also note that the relaxation time for $\tau = 0.4$ is of the order of 
$10^3 \sim 10^4$ times that for $\tau < 0.2$. 
The differences of hysteresis loops and relaxation times 
seem to suggest an existence of a new phase. 
However, there exists another possibility: 
the stationary state indicated by this hysteresis loop 
is artificial due to finite size effects and a new phase does not exist. 
The congestion pattern changes continuously around $\tau \sim 0.22$ (sec) 
and we cannot find a definite transition point. 
It is a future work whether this pattern indicates the existence of 
new phase or not.

\section{Summary and Discussion}

In this paper we investigated the properties of Optimal Velocity Model
with the explicit delay of driver's response. The effects of 
the explicit delay are very small, if the delay time is small:
$\tau < 0.2$ (sec). The effects are similar to the change (reduction) 
of the sensitivity $a$, and therefore the explicit delay does not 
play an essential role. This fact should be compared to the 
traditional car-following models, in which the delay of driver's response 
has played a significant role. The equation of motion of traditional 
car-following model becomes trivial, if the delay time is zero. 

For large explicit delay time $\tau$, the traffic flow behaves in a 
different manner. If $\tau < 0.2$ (sec), the properties of 
congestion clusters are similar to that for $\tau=0$.
For $\tau > 0.2$ (sec), 
the stationary pattern of the traffic flow does not consist of 
only such congestion clusters but confused patterns. 
For $\tau > 0.3$ (sec), the traffic flow becomes stationary but 
congestion clusters are never formed.

In OVM, there is an upper bound of the explicit delay time,
which comes from the condition that the equation of motion 
is meaningful. The upper bound, however, becomes small, if 
we require the existence of stable congestion clusters. 

From this work, we can obtain an indication on a phenomenological study.
In this paper we clarified the notions of the delay time $\tau$ of driver's 
response and the delay time $T$ of car motion.
However the meaning of the delay time of response and 
its effect are model-dependent. 
In traditional car-following models, the delay time $\tau$ seems to 
be merely a fitting parameter and so we can take any value for $\tau$. 
Moreover, the delay time often takes different value in each term. 
In OVM, the delay time $\tau$ is not free and the observed 
value decided by experiments will give a criterion whether 
OVM with OVF (\ref{OVreal}) is valid or not. 
Here we note that the contribution of the delay of driver's 
response to the delay of car motion is very small.
The delay of car motion, therefore, has it's root just in the 
dynamical equation itself. 
This fact suggests the difficulty to determine the delay time $\tau$ of 
driver's response by measuring the delay time $T$ of car motion.
Therefore $\tau$ must be measured directly by other experiments.

\acknowledgments
The authors are grateful to Y.~Sugiyama for helpful discussions.

\newpage

\newpage

\begin{table}[ht]
\begin{tabular}{ccccc}
$\Delta x$ (m) & $f^{-1}$ (s)
& $T_{\tau=0}$ (s) & $T_{\tau=0.1}$ (s) & $T_{\tau=0.2}$ (s) \\ \hline
10 & 2.6427 & 2.6  & 2.6  & 2.6  \\
15 & 1.3434 & 1.35 & 1.35 & 1.35 \\
20 & 0.8282 & 0.95 & 0.95 & 0.95 \\
25 & 0.6921 & 0.85 & 0.87 & 0.89 \\
30 & 0.8282 & 0.95 & 0.95 & 0.95 \\
35 & 1.3434 & 1.35 & 1.35 & 1.35 \\
40 & 2.6427 & 2.6  & 2.6  & 2.6  \\
50 & 13.101 & 13   & 13   & 13   \\
\end{tabular}
\caption{Delay times of car motions in homogeneous flows.}
\label{Tab1}
\end{table}

\begin{table}[ht]
\begin{tabular}{cc}
 $\tau$ & $T$ simulation \\ \hline
 0.0 & 0.94  \\
 0.1 & 0.96  \\
 0.2 & 0.99  \\
\end{tabular}
\caption{Delay times of car motions in congested flows.}
\label{Tab2}
\end{table}

\begin{table}[ht]
\begin{tabular}{ccc}
$\tau$ (s) & $T$ for headway 7m (s) & $T$ for headway 3m (s) \\ \hline
 0.0 & 1.10 & 1.26 \\
 0.1 & 1.10 & 1.26 \\
 0.2 & 1.11 & 1.25 \\
 0.3 & 1.12 & 1.26 \\
\end{tabular}
\caption{Delay times of car motions in queues starting from a traffic signal.}
\label{Tab3}
\end{table}

\end{document}